\documentclass[10pt]{article}
\usepackage[german, english]{babel}
\usepackage[a4paper, left=2.5cm, right=2.5cm, top=3.5cm, bottom=3cm]{geometry}
\usepackage{amssymb}
\usepackage{amsmath}
\usepackage{bbm}
\usepackage{simplewick}
\usepackage{graphicx}
\usepackage[hidelinks]{hyperref}
\usepackage{dsfont}
\usepackage{mathrsfs}
\usepackage{tikz-feynman}
\usepackage{float}
\usepackage{xcolor}
\hypersetup{
 colorlinks=true,
 linkcolor=blue,
 citecolor=magenta,
}

\newcommand{\Z}{\mathds Z}

\newcommand{\N}{\mathds N}
\newcommand{\R}{\mathds R}

\newcommand{\rR}{\mathrm{R}}
\newcommand{\cG}{\mathcal{G}}

\newcommand{\Sl}{\stackrel{\leftarrow}{S}}
\newcommand{\Sr}{\stackrel{\rightarrow}{S}}

\def\im{\mathrm{i}}
\def\ep{\mathrm{e}}
\def\pa{\partial}
\def\diff{\mathrm{d}}

\def\sfrac#1#2{{\textstyle\frac{#1}{#2}}}
\def\>{\rangle}
\def\<{\langle}
\def\||{|\!|}
\def\+{\dagger}

\def\={\ =\ }

\def\und{\quad\textrm{and}\quad}
\def\with{\quad\textrm{with}\quad}

\begin{document}

\title{\bf\huge Supersymmetric large-order perturbation \\[4pt] with the Nicolai map}
\date{~}

\author{\phantom{.}\\[12pt]
{\scshape\Large Olaf Lechtenfeld}
\\[24pt]
Institut f\"ur Theoretische Physik\\ and\\ 
Riemann Center for Geometry and Physics\\[8pt]
Leibniz Universit\"at Hannover \\ 
Appelstra{\ss}e 2, 30167 Hannover, Germany
\\[24pt]
} 

\clearpage
\maketitle
\thispagestyle{empty}

\begin{abstract}
\noindent\large
In rigidly supersymmetric quantum theories, the Nicolai map allows one to turn on a coupling constant 
(from zero to a finite value) by keeping the (free) functional integration measure but subjecting 
the fields to a particular nonlocal and nonlinear transformation. A recursive perturbative construction 
of the Nicolai-transformed field configuration expresses it as a power series in the coupling, 
with its coefficient function at order~$n$ being a sum of particular tree diagrams. 
For a quantum-mechanical example, the size of these tree diagrams (under a certain functional norm) 
is estimated by the $(n{+}1)$st power of the field size, and their number grows like
$n^{-3/2}\times4.967^{\;n}$. Such an asymptotic behavior translates to a finite convergence radius for 
the formal perturbative expansion of the Nicolai map, which establishes its non-perturbative existence.  
The known factorial growth of the number of Feynman diagrams for quantum correlators is reproduced by 
the combinatorics of free-field Wick contractions as usual. 
We expect our results to extend to higher dimensions, including super Yang--Mills theory.
\end{abstract}

\newpage
\setcounter{page}{1} 

\noindent
{\bf Introduction.\ }
Supersymmetry improves the ultraviolet behavior of quantum field theory. 
Technically, this is realized by systematic cancellations of the leading UV divergences
between boson and fermion loops in Feynman diagrams. 
This feature has been exploited systematically by reshuffling perturbation theory
in terms of the inverse Nicolai map~$T_g^{-1}$ via
\begin{equation} \label{globalflow}
\bigl\< X[\phi] \bigr\>_g \= \bigl\< X[T_g^{-1}\phi] \bigr\>_0
\qquad\forall\,X \ ,
\end{equation}
where $\phi$ denotes a collection of bosonic fields, $g$ is a coupling constant,
and the correlators (at coupling~$g$) are taken in an off-shell (rigid) supersymmetric theory 
after integrating out anticommuting and auxiliary degrees of freedom.
The right-hand side inserts a tree expansion (in powers of~$g$) into the functional~$X$
and performs free-boson ($g{=}0$) contractions among all the leaves~$\phi$ of the ensuing tree diagrams.
Since the tree branches are just chains of free fermion propagators,
the resulting `Nicolai rules' thereby remove any fermion loop from the perturbative expansion of~$\<X\>_g$.
It has been demonstrated that this alternative technique is comparable in effort to the conventional Feynman one.
For an early review, see~\cite{Niclectures}. Recent advances on the Nicolai map are \cite{ANPP}--\cite{LR3}.

In an attempt to gain nonperturbative information it is revealing to study the behavior 
of perturbation theory in large order.
One may therefore wonder whether the Nicolai-map technique can also restructure 
the asymptotic properties of supersymmetric perturbative expansions.
In  fact, this question comes in two parts. 
Firstly, one should elucidate the perturbative construction of the map $\phi\mapsto T_g\phi$
(and its inverse) in powers of~$g$, i.e.~the convergence properties of its tree expansion.
Secondly, one must analyze the magnitude of the free-boson contractions~$\<\ldots\>_0$
on the right-hand side of~\eqref{globalflow}.
In this two-stage process, the second stage reduces to the well-known free-field combinatorics of Wick's theorem.
On the other hand, the first stage encodes the entire perturbative expansion of the interacting theory.
This vision was already raised in 1982~\cite{Nic82}.

To address the convergence properties of the perturbative construction of the Nicolai map, 
we must go beyond its formal power-series definition and focus not only on the proliferation 
of tree diagrams with increasing power of~$g$, but also on the `size' of each such diagram.
Since $T_g$ is a map in a certain field space, we should be able to compare its image to its pre-image
under a suitable functional norm. Furthermore, the Nicolai-map and Feynman techniques still share 
the evaluation of gamma-matrix traces (along the tree branches and around the fermion loops, respectively),
which adds to the combinatorial complexity in roughly the same amounts.
Both issues are considerably simpler in $0{+}1$ spacetime dimensions, i.e.~for quantum mechanical models.

\noindent
{\bf Nicolai map for supersymmetric quantum mechanics.\ }
In this letter, we therefore study the large-order behavior of a Nicolai map for the test case of
supersymmetric quantum mechanics with a single bosonic particle coordinate~$x(t)$ 
and velocity $\dot{x}(t)\equiv\tfrac{\diff}{\diff t} x(t)$, a pair $\bigl(\psi(t),\bar\psi(t)\bigr)$ of Grassmann coordinates, 
and a superpotential
\begin{equation}
V(x) \= \tfrac12 mx^2 + \tfrac13 gx^3
\end{equation}
as well as a topological theta term~$\im\theta\,\tfrac{\diff}{\diff t}V$ in the Lagrangian
\begin{equation}
{\cal L} \= \tfrac12 \dot x^2 - \tfrac12 m^2x^2-mgx^3-\tfrac12 g^2x^4+
            \bar\psi\bigl[\im\tfrac{\diff}{\diff t}-m-2gx\bigr]\psi+\im\theta\,(mx+gx^2)\,\dot{x}\ .
\end{equation}
We shall keep the parameters $m$ and $\theta$ fixed and investigate a Nicolai map
linking the free theory ($g{=}0$) to the interacting one ($g{>}0$).
For more details, see~\cite{LR3}.
Denoting compositions of $n\in\N$ by a multiindex
\begin{equation}
{\bf n} = (n_1,n_2,\ldots,n_s) \quad\with n_i\in\N \und \sum_i n_i = n\ ,
\end{equation}
the universal formula for the Nicolai map yields a power series in~$g$~\cite{LR1},
\begin{equation} \label{closedT}
T_g(m,\theta)\,x \= \overrightarrow{\cal P} \exp \Bigl\{-g\!\int_0^1\!\diff\lambda\ R_{\lambda g}(m,\theta)\Bigr\}\ x
\= x\ +\ \sum_{\bf n} g^n\,c_{\bf n}\,\rR_{n_s}(m,\theta)\ldots \rR_{n_2}(m,\theta)\,\rR_{n_1}(m,\theta)\ x\ ,
\end{equation}
with coefficients
\begin{equation} \label{cn}
c_{\bf n} \ =\ (-1)^s\bigl[ n_1\cdot(n_1+n_2)\cdots(n_1+n_2+\ldots+n_s)\bigr]^{-1} 
\ \in\ (-1)^s\,\bigl[ \tfrac{1}{n!},\tfrac{1}{n}\bigr]\ .
\end{equation}
Here, $\overrightarrow{\cal P}$ indicates standard path ordering, and the coupling flow operator
\begin{equation} \label{closedR}
R_g(m,\theta) \= \int\!\diff t''\,\diff t'\ x(t'')^2\  
\bigl[ \tfrac{1+\theta}{2} \, S_g(m;t',t'')\ +\ \tfrac{1-\theta}{2} \, S_g(m;t'',t') \bigr]\ \sfrac{\delta}{\delta x(t')} \\
\ =:\  \sum_{k=1}^\infty g^{k-1}\, \rR_{k}(m,\theta) 
\end{equation}
gets his $g$ and $m$ dependence from the fermion propagator~$S_g(m;t,t')$ defined via
\begin{equation}
\bigl[\im\diff_t- m-2g\,x(t)\bigr]\,S_g(m;t,t')\=\delta(t{-}t')\ .
\end{equation}

One must expand
\begin{equation} \label{expandS}
S_g(m;t,t') \= S_0(m;t{-}t')\ +\ 2g\int\!\diff{u}\ S_0(m;t{-}u)\,x(u)\,S_0(m;u{-}t')\ +\ O(g^2)
\end{equation}
to read off the power series in~\eqref{closedR} whose terms enter the series in~\eqref{closedT}.
For the free massive fermion propagator~$S_0(m;t{-}t')$ we may choose the retarded one and 
Wick rotate to Euclidean time, so that
\begin{equation} \label{freeprop}
S_0(m;t{-}t') \= \Theta(t{-}t')\;\ep^{-m(t-t')}
\end{equation}
enters as a building block in~\eqref{closedR}.
It is customary to graphically represent $S_0(m;t{-}t')$ by a solid line connecting vertices 
representing the instances $t$ and~$t'$ and to depict factors of $x(u)$ by a wavy leaf 
attached to~$u$. 
Here is the picture:
\begin{equation}
                \begin{aligned}
                        R_g(m,\theta)\=&\Bigl\{
                        \tfrac{1{+}\theta}{2}\begin{tikzpicture}[baseline={([yshift=-.1cm]current bounding box.center)}, scale=0.4, transform shape]
                                \begin{feynman}
                                        \vertex (a);
                                        \vertex [right=of a] (c);
                                        \vertex [above left=of a] (d);
                                        \vertex [below left=of a] (e);
                                        \diagram* {
                                                (a) -> [fermion] (c),
                                                (a) -- [photon] (d),
                                                (a) -- [photon] (e),
                                        };
                                \end{feynman}\;\end{tikzpicture}
                        \ +\ \tfrac{1{-}\theta}{2}
                        \begin{tikzpicture}[baseline={([yshift=-.1cm]current bounding box.center)}, scale=0.4, transform shape]
                                \begin{feynman}
                                        \vertex (a);
                                        \vertex [right=of a] (c);
                                        \vertex [above left=of a] (d);
                                        \vertex [below left=of a] (e);
                                        \diagram* {
                                                (a) -> [anti fermion] (c),
                                                (a) -- [photon] (d),
                                                (a) -- [photon] (e),
                                        };
                                \end{feynman}\;\end{tikzpicture}\Bigr\}
                        \ +\ g\ \Bigl\{\tfrac{1{+}\theta}{2}\begin{tikzpicture}[baseline={([yshift=-.16cm]current bounding box.center)}, scale=0.4, transform shape]
                                \begin{feynman}
                                        \vertex (a);
                                        \vertex [right=of a] (b);
                                        \vertex [right=of b] (c);
                                        \vertex [above left=of a] (d);
                                        \vertex [below left=of a] (e);
                                        \vertex [above=of b] (f);
                                        \diagram* {
                                                (a) -- [fermion] (b) -> [fermion] (c),
                                                (a) -- [photon] (d),
                                                (b) -- [photon] (f),
                                                (a) -- [photon] (e),
                                        };
                                \end{feynman}\;\end{tikzpicture}
                        \ +\ \tfrac{1{-}\theta}{2}
                        \begin{tikzpicture}[baseline={([yshift=-.16cm]current bounding box.center)}, scale=0.4, transform shape]
                                \begin{feynman}
                                        \vertex (a);
                                        \vertex [right=of a] (b);
                                        \vertex [right=of b] (c);
                                        \vertex [above left=of a] (d);
                                        \vertex [below left=of a] (e);
                                        \vertex [above=of b] (f);
                                        \diagram* {
                                                (a) -- [anti fermion] (b) -> [anti fermion] (c),
                                                (a) -- [photon] (d),
                                                (b) -- [photon] (f),
                                                (a) -- [photon] (e),
                                        };
                                \end{feynman}\;\end{tikzpicture}\Bigr\}\\
                        +\ &g^2\ \Bigl\{\tfrac{1{+}\theta}{2}
                        \begin{tikzpicture}[baseline={([yshift=-.16cm]current bounding box.center)}, scale=0.4, transform shape]
                                \begin{feynman}
                                        \vertex (a);
                                        \vertex [right=of a] (b);
                                        \vertex [right=of b] (c);
                                        \vertex [right=of c] (d);
                                        \vertex [above left=of a] (o1);
                                        \vertex [below left=of a] (o2);
                                        \vertex [above=of b] (be);
                                        \vertex [above=of c] (ce);
                                        \diagram* {
                                                (a) -- [fermion] (b) -- [fermion] (c) -> [fermion] (d),
                                                (b) -- [photon] (be),
                                                (c) -- [photon] (ce),
                                                (a) -- [photon] (o1),
                                                (a) -- [photon] (o2),
                                        };
                                \end{feynman}\;
                        \end{tikzpicture}\ +\ \tfrac{1{-}\theta}{2}
                        \begin{tikzpicture}[baseline={([yshift=-.16cm]current bounding box.center)}, scale=0.4, transform shape]
                                \begin{feynman}
                                        \vertex (a);
                                        \vertex [right=of a] (b);
                                        \vertex [right=of b] (c);
                                        \vertex [right=of c] (d);
                                        \vertex [above left=of a] (o1);
                                        \vertex [below left=of a] (o2);
                                        \vertex [above=of b] (be);
                                        \vertex [above=of c] (ce);
                                        \diagram* {
                                                (a) -- [anti fermion] (b) -- [anti fermion] (c) -> [anti fermion] (d),
                                                (b) -- [photon] (be),
                                                (c) -- [photon] (ce),
                                                (a) -- [photon] (o1),
                                                (a) -- [photon] (o2),
                                        };
                                \end{feynman}\;
                        \end{tikzpicture}\Bigr\}\ +\ \mathcal{O}(g^4)\ ,
                \end{aligned}
        \end{equation}
where the arrows at the end of the fermion lines indicate a functional derivative with respect to $x$.
For the precise graphical rules see~\cite{LR3}.
In this way, $\rR_k$ acts derivatively on functionals of~$x$ by successively
replacing each `leaf'~$x(t)$ with a tree of `height'~$k$ (the number of $S_0$ factors)
rooted at~$t$ and containing $k{+}1$ leaves of its own:
a leaf~$x(u)$ at each $S_0$ juncture and a double leaf~$x(t'')^2$ at the tree top.
All times except at the root are integrated over.
Ignoring the leaves, we have a propagator `skeleton' which is unbranched.

The repeated action of $\rR_{n_i}$ on each other in~\eqref{closedT} then
successively grafts $s$~such trees on each other, creating also branched skeletons.
More explicitly, grafting at the top of a tree simply extends it 
while grafting at the stem creates a new branch extending from a then leafless vertex.
In this fashion, the alternating infinite sum~\eqref{closedT} generates 
all possible (unordered) rooted trees of a certain kind,
with $n$~vertices (excluding the root), $n$ free fermion lines, and $n{+}1$ leaves.
If we include the leaves into the set of $2n{+}1$ nodes, then such trees are 
named `strictly binary' because every node has exactly two (in case of a vertex)
or zero (in case of a leaf) children. They are also known as `Otter trees'.
If we strip off the leaves and consider only the skeleton, then such trees are
called `weakly binary' since each vertex has at most two children,
namely zero for a branch terminal, two for a branch point, and one otherwise.
The number of distinct such trees without labelling is the ($n{+}1$)st
`Wedderburn--Etherington number' WE$(n{+}1)$.

Finally, the tree diagrams carry certain weights $w[\mathrm{tree}]$,
and each of the fermion lines is oriented.
The flow operator~\eqref{closedR} is a sum of two terms,
the first one containing a `top-directed' full propagator (weighted with $\tfrac{1{+}\theta}{2}$),
and the second one a `root-directed' full propagator (weighted with $\tfrac{1{-}\theta}{2}$).
From~\eqref{expandS} it follows that this orientation is uniform from root to top
along any of the two unbranched skeletons from~$\rR_k$.
The grafting process, however, produces trees with any distribution of orientations
for the free fermion propagators making up their skeleton.
This is already clear from the $\rR_1^n$ term at order $g^n$ in~\eqref{closedT}.
Since a tree with a particular `orientation dressing' usually appears in more than one
grafting sequence encoded by~${\bf n}$,
its weight $w[\mathrm{tree}](\theta)$ is a {\it sum\/} of products of $s$~factors of
$\tfrac{1{+}\theta}{2}$ or $\tfrac{1{-}\theta}{2}$
multiplied by the corresponding $c_{\bf n}$ coefficients~\eqref{cn}  in~\eqref{closedT}.
The result is $1/(2n)!!$ times a polynomial of order~$n$ in~$\theta$ with integral coefficients.
It is convenient to absorb the factors of 2 for each non-terminal vertex (see~\eqref{expandS})
into the graphical `Nicolai rules' for the vertices, just like the integrations over all times but the root one.
We display the tree expansion with proper weights up to order~$g^3$:
\begin{equation}\label{eq:map_interacting_theta}
                \begin{aligned}
                        T_g(m,\theta)\ x(\omega)\=& x(\omega)
                        \ -\ \sfrac{g}{2}\ \Bigl\{
                        (1{+}\theta)\begin{tikzpicture}[baseline={([yshift=-.1cm]current bounding box.center)}, scale=0.4, transform shape]
                                        \begin{feynman}
                                                \vertex (a);
                                                \vertex [right=of a] (c);
                                                \vertex [above left=of a] (d);
                                                \vertex [below left=of a] (e);
                                                \diagram* {
                                                        (a) -- [fermion] (c),
                                                        (a) -- [photon] (d),
                                                        (a) -- [photon] (e),
                                                };
                                        \end{feynman}\;\end{tikzpicture}
                        \ +\ (1{-}\theta)
                        \begin{tikzpicture}[baseline={([yshift=-.1cm]current bounding box.center)}, scale=0.4, transform shape]
                                        \begin{feynman}
                                                \vertex (a);
                                                \vertex [right=of a] (c);
                                                \vertex [above left=of a] (d);
                                                \vertex [below left=of a] (e);
                                                \diagram* {
                                                        (c) -- [fermion] (a),
                                                        (a) -- [photon] (d),
                                                        (a) -- [photon] (e),
                                                };
                                        \end{feynman}\;\end{tikzpicture}
                        \Bigr\}\\ \ -\ & \sfrac{g^2}{8}\ (1{+}\theta)(1{-}\theta)\Bigl\{\begin{tikzpicture}[baseline={([yshift=-.16cm]current bounding box.center)}, scale=0.4, transform shape]
                                        \begin{feynman}
                                                \vertex (a);
                                                \vertex [right=of a] (b);
                                                \vertex [right=of b] (c);
                                                \vertex [above left=of a] (d);
                                                \vertex [below left=of a] (e);
                                                \vertex [above=of b] (f);
                                                \diagram* {
                                                        (a) -- [fermion] (b) -- [fermion] (c),
                                                        (b) -- [photon] (f),
                                                        (a) -- [photon] (d),
                                                        (a) -- [photon] (e),
                                                };
                                        \end{feynman}\;
                        \end{tikzpicture}\ +\ \begin{tikzpicture}[baseline={([yshift=-.16cm]current bounding box.center)}, scale=0.4, transform shape]
                                        \begin{feynman}
                                                \vertex (a);
                                                \vertex [right=of a] (b);
                                                \vertex [right=of b] (c);
                                                \vertex [above left=of a] (d);
                                                \vertex [below left=of a] (e);
                                                \vertex [above=of b] (f);
                                                \diagram* {
                                                        (c) -- [fermion] (b) -- [fermion] (a),
                                                        (b) -- [photon] (f),
                                                        (a) -- [photon] (d),
                                                        (a) -- [photon] (e),
                                                };
                                        \end{feynman}\;
                        \end{tikzpicture}\ -\ \begin{tikzpicture}[baseline={([yshift=-.16cm]current bounding box.center)}, scale=0.4, transform shape]
                                        \begin{feynman}
                                                \vertex (a);
                                                \vertex [right=of a] (b);
                                                \vertex [right=of b] (c);
                                                \vertex [above left=of a] (d);
                                                \vertex [below left=of a] (e);
                                                \vertex [above=of b] (f);
                                                \diagram* {
                                                        (a) -- [fermion] (b),
                                                        (c) -- [fermion] (b),
                                                        (b) -- [photon] (f),
                                                        (a) -- [photon] (d),
                                                        (a) -- [photon] (e),
                                                };
                                        \end{feynman}\;
                        \end{tikzpicture}\ -\ \begin{tikzpicture}[baseline={([yshift=-.16cm]current bounding box.center)}, scale=0.4, transform shape]
                                        \begin{feynman}
                                                \vertex (a);
                                                \vertex [right=of a] (b);
                                                \vertex [right=of b] (c);
                                                \vertex [above left=of a] (d);
                                                \vertex [below left=of a] (e);
                                                \vertex [above=of b] (f);
                                                \diagram* {
                                                        (b) -- [fermion] (a),
                                                        (b) -- [fermion] (c),
                                                        (b) -- [photon] (f),
                                                        (a) -- [photon] (d),
                                                        (a) -- [photon] (e),
                                                };
                                        \end{feynman}\;
                        \end{tikzpicture}\Bigr\}\\ \ -\ &\sfrac{g^3}{48}(1{+}\theta)(1{-}\theta)\Bigl\{(3{-}\theta)\begin{tikzpicture}[baseline={([yshift=-.16cm]current bounding box.center)}, scale=0.4, transform shape]
                                        \begin{feynman}
                                                \vertex (a);
                                                \vertex [right=of a] (b);
                                                \vertex [right=of b] (c);
                                                \vertex [right=of c] (d);
                                                \vertex [above left=of a] (o1);
                                                \vertex [below left=of a] (o2);
                                                \vertex [above=of b] (be);
                                                \vertex [above=of c] (ce);
                                                \diagram* {
                                                        (a) -- [fermion] (b) -- [fermion] (c) -- [fermion] (d),
                                                        (b) -- [photon] (be),
                                                        (c) -- [photon] (ce),
                                                        (a) -- [photon] (o1),
                                                        (a) -- [photon] (o2),
                                                };
                                       \end{feynman}\;
                        \end{tikzpicture}
                \ -\ (1{+}\theta)
                \begin{tikzpicture}[baseline={([yshift=-.4cm]current bounding box.center)}, scale=0.4, transform shape]
                \begin{feynman}
                \vertex (a);
                \vertex [right=of a] (b);
                \vertex [right=of b] (c);
                \vertex [above left=of a] (d);
                \vertex [below left=of a] (e);
                \vertex [above=of b] (f);
                \vertex [above right=of f] (f1);
                \vertex [above left=of f] (f2);
                \diagram* {
                        (a) -- [fermion] (b) -- [fermion] (c),
                        (b) -- [anti fermion] (f),
                        (f) -- [photon] (f1),
                        (f) -- [photon] (f2),
                        (a) -- [photon] (d),
                        (a) -- [photon] (e),
                };
                \end{feynman}\;
                \end{tikzpicture}\\
                &\ \ -\         (1{-}\theta)\begin{tikzpicture}[baseline={([yshift=-.16cm]current bounding box.center)}, scale=0.4, transform shape]
                                        \begin{feynman}
                                                \vertex (a);
                                                \vertex [right=of a] (b);
                                                \vertex [right=of b] (c);
                                                \vertex [right=of c] (d);
                                                \vertex [above left=of a] (o1);
                                                \vertex [below left=of a] (o2);
                                                \vertex [above=of b] (be);
                                                \vertex [above=of c] (ce);
                                                \diagram* {
                                                        (a) -- [anti fermion] (b) -- [fermion] (c) -- [fermion] (d),
                                                        (b) -- [photon] (be),
                                                        (c) -- [photon] (ce),
                                                        (a) -- [photon] (o1),
                                                        (a) -- [photon] (o2),
                                                };
                                        \end{feynman}\;
                        \end{tikzpicture}
                        \ -\ (3{+}\theta)
                        \begin{tikzpicture}[baseline={([yshift=-.16cm]current bounding box.center)}, scale=0.4, transform shape]
                                        \begin{feynman}
                                                \vertex (a);
                                                \vertex [right=of a] (b);
                                                \vertex [right=of b] (c);
                                                \vertex [right=of c] (d);
                                                \vertex [above left=of a] (o1);
                                                \vertex [below left=of a] (o2);
                                                \vertex [above=of b] (be);
                                                \vertex [above=of c] (ce);
                                                \diagram* {
                                                        (a) -- [anti fermion] (b) -- [anti fermion] (c) -- [fermion] (d),
                                                        (b) -- [photon] (be),
                                                        (c) -- [photon] (ce),
                                                        (a) -- [photon] (o1),
                                                        (a) -- [photon] (o2),
                                                };
                                        \end{feynman}\;
                        \end{tikzpicture}
                        \ +\ (1{+}\theta)
                        \begin{tikzpicture}[baseline={([yshift=-.16cm]current bounding box.center)}, scale=0.4, transform shape]
                                        \begin{feynman}
                                                \vertex (a);
                                                \vertex [right=of a] (b);
                                                \vertex [right=of b] (c);
                                                \vertex [right=of c] (d);
                                                \vertex [above left=of a] (o1);
                                                \vertex [below left=of a] (o2);
                                                \vertex [above=of b] (be);
                                                \vertex [above=of c] (ce);
                                                \diagram* {
                                                        (a) -- [fermion] (b) -- [anti fermion] (c) -- [fermion] (d),
                                                        (b) -- [photon] (be),
                                                        (c) -- [photon] (ce),
                                                        (a) -- [photon] (o1),
                                                        (a) -- [photon] (o2),
                                                };
                                        \end{feynman}\;
                        \end{tikzpicture}\\
                        &\ \ +\ 2\theta
                        \begin{tikzpicture}[baseline={([yshift=-.4cm]current bounding box.center)}, scale=0.4, transform shape]
                                        \begin{feynman}
                                                \vertex (a);
                                                \vertex [right=of a] (b);
                                                \vertex [right=of b] (c);
                                                \vertex [above left=of a] (d);
                                                \vertex [below left=of a] (e);
                                                \vertex [above=of b] (f);
                                                \vertex [above right=of f] (f1);
                                                \vertex [above left=of f] (f2);
                                                \diagram* {
                                                        (a) -- [fermion] (b) -- [fermion] (c),
                                                        (b) -- [fermion] (f),
                                                        (f) -- [photon] (f1),
                                                        (f) -- [photon] (f2),
                                                        (a) -- [photon] (d),
                                                        (a) -- [photon] (e),
                                                };
                                        \end{feynman}\;
                        \end{tikzpicture}
                        \ +\ (1{-}\theta) \begin{tikzpicture}[baseline={([yshift=-.4cm]current bounding box.center)}, scale=0.4, transform shape]
                                        \begin{feynman}
                                                \vertex (a);
                                                \vertex [right=of a] (b);
                                                \vertex [right=of b] (c);
                                                \vertex [above left=of a] (d);
                                                \vertex [below left=of a] (e);
                                                \vertex [above=of b] (f);
                                                \vertex [above right=of f] (f1);
                                                \vertex [above left=of f] (f2);
                                                \diagram* {
                                                        (a) -- [anti fermion] (b) -- [fermion] (c),
                                                        (b) -- [fermion] (f),
                                                        (f) -- [photon] (f1),
                                                        (f) -- [photon] (f2),
                                                        (a) -- [photon] (d),
                                                        (a) -- [photon] (e),
                                                };
                                        \end{feynman}\;
                        \end{tikzpicture}
                        \ +\ \bigl(\theta\leftrightarrow-\theta,\Sr_0\leftrightarrow\Sl_0\bigr)\Bigr\}\ +\ \mathcal{O}(g^4)\ ,
                \end{aligned}
        \end{equation}
where $\Sr_0\leftrightarrow\Sl_0$ indicates a reversal of the orientation of each fermion propagator.
Note that the first diagram in the last line appears in two ways (permuted by the two branches),
which yields the `2' in the weight.
For the count of orientation-dressed graphs we thus have to consider $\Z_2$ labelled trees.
Multiplying the previous WE$(n{+}1)$ by a factor of~$2^n$ only slightly overcounts their number,
since like branches with different orientations may be related by permutation symmetry,
but this is compensated by the weight factors.

The inverse Nicolai map has an analogous representation, which is obtained from~\eqref{closedT}
by reversing the path ordering and the sign in the exponent. 
Its graphical representation can be found in~\cite{LR3}.
The effect on the power series is to replace the coefficients $c_{\bf n}$ with~\cite{LR1}
\begin{equation}
d_{\bf n} \ = \bigl[ n_s\cdot(n_s+n_{s-1})\cdots(n_s+n_{s-1}+\ldots+n_1)\bigr]^{-1}
\ \in\ \bigl[ \tfrac{1}{n!},\tfrac{1}{n}\bigr]\\ ,
\end{equation}
so it is no longer alternating.

\noindent
{\bf Estimating the size of a tree diagram.\ }
Let us consider a generic tree graph~$\cG_{n,b}(t)$ with $n$~free fermion propagators 
in $b$~branches and rooted at time~$t$. We also fix a distribution of propagator orientations.
Then, $b$ of its $n$~vertices are terminal (with two leaves attached), $b{-}1$ of them are branch points
(without a leaf), and the remaining $n{-}2b{+}1$ are stem vertices (carrying a single leaf).
We like to estimate firstly a suitable norm of $\cG_{n,b}(t)$ as a function of the norm 
of the functional variable~$x(t)$ and secondly the weight~$w[\cG_{n,b}](\theta)$.
For the first goal we should try to bound the graph by a product of vertex contributions.
According to~\eqref{freeprop}, $S_0$ is bounded from above by the constant function~$1$,
so we can estimate by unity all free-fermion lines of our graph, except for the one
connected to the root.
Then, for the different vertices we are left with factors less than
\begin{equation}
\begin{aligned}
\smallint \!\diff u\  \bigl| x(u)^2 \bigr| &\= \|| x^2 \||_1 \qquad\;\textrm{for a terminal vertex}\ , \\
2\smallint \!\diff u\ \bigl| x(u)  \bigr| &\= 2\, \|| x \||_1 \qquad\textrm{for a stem vertex}\ ,\\[2pt]
2\smallint \!\diff u\ 1 &\= 2\, \|| 1 \||_1 \qquad\,\textrm{for a branchpoint vertex}\ ,
\end{aligned}
\end{equation}
where we introduced the $L^p$ norm
\begin{equation}
\|| f \||_p \= \biggl( \int\!\diff u\ \bigl| f(u) \bigr|^p \biggr)^{1/p}\ .
\end{equation}
Integrating $u$ over the whole real axis lets the branchpoint contribution diverge, 
but in this case the function~$x(u)$ is absent so we can do better by explicitly integrating
\begin{equation}
2\int_D\!\diff u\ \ep^{\epsilon_1 m(t_1-u)}\ \ep^{\epsilon_2 m(t_2-u)}\ \ep^{\epsilon_3 m(u-t_3)} \ \le\ \frac2m\ ,
\end{equation}
where the $\epsilon_i\in\{+1,-1\}$ encode the orientation of the three attached lines, 
and $D$ is the domain where all exponents are nonpositive.
If the first vertex attached to the root is not a branch point, we then obtain that
\begin{equation}
\bigl| \cG_{n,b}(t) \bigr| \ \le\ 
\bigl(\|| x^2 \||_1\bigr)^b\ \bigl(\tfrac2m\bigr)^{b-1}\ \bigl( 2\, \|| x \||_1 \bigr)^{n-2b}\ 
2\int_D\!\diff u\ \bigl| x(u) \bigr|\ \ep^{\epsilon m(u-t)}
\end{equation}
with $\epsilon$ and $D$ as above. It follows that
\begin{equation} \label{norm1G}
\bigl|\!\bigl| \cG_{n,b} \bigr|\!\bigr|_1 \ \le\ 2^{n-b}\,m^{-b}\ \|| x \||_1^{n-2b+1}\ \|| x^2 \||_1^b\ .
\end{equation}
If the first vertex attached to the root is a branch point, then the $L^1$ norm diverges 
since the $u$~integral equals $\frac1m$ independent of~$t$.
A more careful consideration retaining also the next propagator into the branch shows, however, that
the same bound can be achieved for this case.

A drawback of \eqref{norm1G} is that we need the norm of $x^2$, which is not directly related to the one of~$x$ itself.
When replacing free propagators by unity, we partially forfeited the infrared suppression given by the finite mass~$m$.
This can be repaired by replacing the infinite time integration domain by a finite one, say of length~$1/m$,
which acts as an infrared regulator and renders the unit function measurable. 
Equivalently, we could demand $x(t)$ to be of finite support on~$\R$. With such a cut-off,
H\"older's inequality implies that
\begin{equation}
\|| f \||_p \ \le\ m^{\frac1q-\frac1p}\ \|| f \||_q \qquad\textrm{for}\quad 1\le p<q\ .
\end{equation}
Since $\|| x^2 \||_1=\|| x \||_2^2$, we better pass to the $L^2$ norm and use the bound 
$\||x\||_1\le\frac1{\sqrt{m}}\||x\||_2$ provided by our infrared cut-off.
In this way we arrive at
\begin{equation}
\bigl| \cG_{n,b}(t) \bigr| \ \le\ 2^{n-b}\,m^{1-\frac{n}{2}}\ \||x\||_2^n\ 
\int_D\!\diff u\ \bigl| x(u) \bigr|\ \ep^{\epsilon m(u-t)}\ ,
\end{equation}
focusing on the (more common) case of a stem vertex next to the root.
Inserting this expression into the $L^2$ norm, we obtain (without further infrared regulators)
\begin{equation}
\begin{aligned}
\bigl|\!\bigl| \cG_{n,b} \bigr|\!\bigr|_2^2 &\ \le\ 
2^{2n-2b}\,m^{2-n}\ \||x\||_2^{2n}\ \int_{\R}\!\diff t \int_D\!\diff u \int_{D'}\!\diff u'\ 
\bigl| x(u) \bigr|\  \bigl| x(u') \bigr|\ \ep^{\epsilon m(u-t)}\ \ep^{\epsilon m(u'-t)} \\
&\= 2^{2n-2b}\,m^{2-n}\ \||x\||_2^{2n}\ \int_{\R}\!\diff t \int_0^\infty\!\!\!\diff u \int_0^\infty\!\!\!\diff u'\ 
\bigl| x(t{-}\epsilon u) \bigr|\  \bigl| x(t{-}\epsilon u') \bigr|\ \ep^{-m(u+u')} \\
&\= 2^{2n-2b}\,m^{2-n}\ \||x\||_2^{2n}\ \tfrac12\int_0^\infty\!\diff(u{+}u')\ \ep^{-m(u+u')} 
\int_{\R}\!\diff t' \int_{-(u+u')}^{+(u+u')}\!\!\!\!\!\!\!\!\!\!\!\!\!\!\!\diff(u{-}u')\ 
\bigl| x(t') \bigr|\  \bigl| x(t'{+}\epsilon(u{-}u')) \bigr| \\[4pt]
&\ \le\ 2^{2n-2b}\,m^{2-n}\ \||x\||_2^{2n}\ \tfrac1{2m}\ \||x\||_1^2
\ \ \le\ \ 2^{2n-2b}\,m^{2-n}\ \||x\||_2^{2n}\ \tfrac1{2m^2}\ \||x\||_2^2\ ,
\end{aligned}
\end{equation}
thus ending with
\begin{equation} \label{Gnorm}
\bigl|\!\bigl| \cG_{n,b} \bigr|\!\bigr|_2 \ \le\ 
2^{n-b-\frac12}\,m^{-\frac{n}{2}}\ \||x\||_2^{n+1} \ \le\
\tfrac1{\sqrt{2}}\ (2/\sqrt{m})^n\ \||x\||_2^{n+1} \ .
\end{equation}
As before, graphs with a branch point immediately above the root yield the same bound.

For the second goal, approximating the weight~$w[\cG_{n,b}](\theta)$,
we recall that, at the `magical values' $\theta={\pm}1$, 
all weights for~$T_g(m,\theta)\,x$ vanish for $n>1$, 
and those for $T^{-1}_g(m,\theta)\,x$ are unity divided by a symmetry factor.
Let us estimate the weight for a generic tree contributing to~$T_g(m,\theta)\,x$.
It is easy to see that the largest weight at order~$g^n$ in the power series~\eqref{closedT} 
belongs to the two ($b{=}1$) trees generated by $\tfrac1n\,\rR_n\,x$.
All its lines have the same orientation.
All grafting sequences contribute to them, and the coefficients add up to
\begin{equation} 
\textrm{max}\,\bigl| w[\cG_{n,b}](\theta)\bigr| \= 
\tfrac{1}{(2n)!!}\,\bigl|(1{\mp}\theta)(1{\pm}\theta)(3{\pm}\theta)\cdots(2n{-}3{\pm}\theta)\bigr|
\ \rightarrow\ \begin{cases}
\tfrac{|\theta|^n}{(2n)!!}\ \sim\ \tfrac{1}{\sqrt{2\pi n}}\bigl|\tfrac{\ep}{2}\theta\bigr|^n n^{-n}
&\textrm{for}\ \theta\to\infty \\[6pt]
\tfrac{(2n-3)!!}{(2n)!!}\ \sim\ \tfrac{1}{\sqrt{\pi n}}\,\tfrac{1}{2n}
&\textrm{for}\ \theta\to 0 \end{cases}
\end{equation}
with the sign depending on the orientation of the entire tree. 
The expressions after the $\sim$ sign describe the large-$n$ asymptotics.
So for $\theta$ of the order of unity, we can safely bound 
$\bigl|w[\cG_{n,b}]\bigr|\le1$.
This bound is supported by the observation that the weights for $T^{-1}_g(m,\theta{\approx}{\pm}1)$
are of the order of unity and cannot vary too wildly with~$\theta$.

\noindent
{\bf Asymptotic growth of the power series.\ }
Combining the considerations above and employing the triangle inequality to
the $L^2$ norm of~\eqref{closedT} we arrive at
\begin{equation} \label{Testimate}
\bigl|\!\bigl| T_g(m,\theta)\,x \bigr|\!\bigr|_2 \ \le\ 
\Bigl( 1\ +\ \sum_{n=1}^\infty g^n \sum_{\cG_{n,b}} \tfrac{1}{\sqrt{2}}\,\bigl( 2\,\||x\||_2/\sqrt{m} \bigr)^n \Bigr)
\ \bigl|\!\bigl|x\bigr|\!\bigr|_2
\qquad\textrm{for}\quad \theta=O(1)\ .
\end{equation}
The $m$~dependence is dictated by dimensional analysis: 
With $g$ of dimension mass${}^{\frac32}$ and $\||x\||_2$ of dimension mass${}^{-1}$,
the $1/\sqrt{m}$ factor makes the large bracket dimensionless.
The inner sum still runs over all orientation-dressed tree graphs with $n{+}1$~leaves.
To extract the asymptotic growth of the power-series coefficients, we thus still have to determine 
the large-order growth of the number of the $\Z_2$-labelled Otter trees. 
The known large-$n$ behavior of the Wedderburn-Etherington numbers yields the count~\footnote{
    The Otter trees proliferate more slowly than the number of ordered (strict) binary trees
    (given by the Catalan numbers).}
\begin{equation}
2^n\,\textrm{WE}(n{+}1) \ \sim\ {0.7916} \times n^{-\frac32} \times {4.967}^{\;n} 
\qquad\textrm{for}\quad n\to\infty\ .
\end{equation}
Inserted into \eqref{Testimate} we finally arrive at
\begin{equation} \label{Testimate2}
\bigl|\!\bigl| T_g(m,\theta)\,x \bigr|\!\bigr|_2 \ \lesssim\ 
\Bigl( 1\ +\ 0.5597 \sum_{n=1}^\infty  n^{-\frac32}\;\bigl( 9.934\; \||x\||_2/\sqrt{m} \bigr)^n\ g^n \Bigr)
\ \bigl|\!\bigl|x\bigr|\!\bigr|_2
\qquad\textrm{for}\quad \theta=O(1)\ .
\end{equation}
Since the coefficient growth is not faster than exponential, 
the power series has a finite radius of convergence given by
$|g|\,\gtrsim\,0.1\,\sqrt{m}\big/\||x\||_2$.
The same should apply to the {\it inverse\/} map $T^{-1}_g x$.

\noindent
{\bf Alternative counting.}
Alternatively, we may treat {\it separately\/} all grafting sequences of order~$g^n$ 
in the sum of~\eqref{closedT}.
Each such summand is uniquely characterized by the composition~${\bf n}$ of~$n$
and an assignment of an orientation to each of its $s$~parts~$n_i$.
Its weight is simply
\begin{equation} \label{nweight}
w({\bf n},\textrm{orientation})\ =\ c_{\bf n}\ \bigl(\tfrac{1+\theta}{2}\bigr)^{s_+} \bigl(\tfrac{1-\theta}{2}\bigr)^{s_-}
\quad\textrm{with}\quad s_+{+}s_-=s\ ,
\end{equation}
where $s_{\pm}$ are the number of top- and root-directed $\rR_{n_i}$ operations in the grafting sequence.
However since the grafting acts as a derivation, any such sequence produces a multitude of different trees. 
Every oriented $\rR_{n_i}$ can act on the $n_1{+}n_2{+}\ldots{+}n_{i-1}+1$
leaves created by the previous grafting actions and thus will multiply the number of trees by at most
$n_1{+}n_2{+}\ldots{+}n_{i-1}$ (as there is at least one two-leaf terminal).
Therefore, an oriented sequence $\rR_{n_s}\cdots\rR_{n_2}\rR_{n_1}\,x$ generates not more than
$n_1{\cdot}(n_1{+}n_2)\cdots(n_1{+}n_2{+}\ldots{+}n_{s-1})$ tree diagrams with some orientation distribution.
Also conversely, a given tree generically arises from several sequences,
but here we simply do not lump them together (which avoids having to add the weights). 
The size of any such diagram can be estimated with~\eqref{Gnorm}.
So we need to bound their numbers weighted with~\eqref{nweight}.
Denoting with ${\bf n}(n)$ the multiindices of a fixed length~$n$, 
taking into account the sum over $2^s$ orientation distributions,
and using $|1{\pm}\theta|\le1{+}|\theta|$,
an upper bound is
\begin{equation}
\begin{aligned}
\sum_{{\bf n}(n)} 2^s\,|c_{\bf n}|\,\bigl(\tfrac{1+|\theta|}{2}\bigr)^s\ 
n_1{\cdot}(n_1{+}n_2)\cdots(n_1{+}n_2{+}\ldots{+}n_{s-1}) &\=
\tfrac1n\,\sum_{s=1}^n {\textstyle\binom{n{-}1}{s{-}1}}\,\bigl(1{+}|\theta|\bigr)^s \\[4pt]
\= \tfrac1n\,\bigl(1{+}|\theta|\bigr) \bigl(2{+}|\theta|\bigr)^{n-1}
&\ \rightarrow\ \begin{cases}
\tfrac1n\,|\theta|^n &\textrm{for}\ \theta\to\infty \\[6pt]
\tfrac1n\,2^{n-1} &\textrm{for}\ \theta\to 0 \end{cases}
\end{aligned}
\end{equation}
where in the first step we used that fact that the number of compositions of~$n$ into $s$~parts
is given by~$\binom{n{-}1}{s{-}1}$, and the second step perfoms the binomial sum.
Carrying over the norm estimate~\eqref{Gnorm}, we collect
\begin{equation} \label{Testimate3}
\bigl|\!\bigl| T_g(m,\theta)\,x \bigr|\!\bigr|_2 \ \lesssim\ 
\Bigl( 1\ +\ \tfrac{1}{\sqrt{8}} \sum_{n=1}^\infty  n^{-1}\;\bigl( 4\, \||x\||_2/\sqrt{m} \bigr)^n\ g^n \Bigr)
\ \bigl|\!\bigl|x\bigr|\!\bigr|_2
\qquad\textrm{for}\quad \theta=O(1)\ .
\end{equation}

\noindent
{\bf Convergence of the Nicolai map expansion.\ }
It is reassuring that both counting methods give the qualitatively similar results
\eqref{Testimate2} and \eqref{Testimate3}, i.e.
\begin{equation}
\bigl|\!\bigl| T_g(m,\theta)\,x \bigr|\!\bigr|_2 \ \lesssim\ 
\Bigl( 1\ +\ \gamma \sum_{n=1}^\infty  n^{-\beta}\;\bigl( \alpha\,\||x\||_2/\sqrt{m} \bigr)^n\ g^n \Bigr)
\ \bigl|\!\bigl|x\bigr|\!\bigr|_2 
\qquad\textrm{for}\quad \theta=O(1)\ ,
\end{equation}
with some constants $\alpha$, $\beta$ and $\gamma$. 
Their numerical values should not be taken very seriously here,
because we have not been too careful with the bounds.
The key issue is the exponential growth of the power-series coefficients, 
which renders the radius of convergence {\it finite\/} (for $m{>}0$ and $\||x\||_2{<}\infty$).
Since the inverse Nicolai map is obtained by a formal power-series inversion, 
this result applies just as well to~$T^{-1}_g\,x$.
Therefore, (at least in quantum mechanics) the expression $X[T_g^{-1}\phi]$ inside the free-theory brackets 
on the right-hand side of~\eqref{globalflow} is not only formally defined, 
but exists {\it beyond the perturbative expansion\/} for sufficiently small values of~$g$
as a functional of the input variable~$\phi$ (and not merely as a function of a spacetime point).
Therefore, $T_g$ and its inverse are well-defined operators in a suitable function space,
which we haven't been too careful in defining it precisely here.
We note that no renormalization is required at this stage, since the trees are classical objects without any loops.

On the other hand, it is known that the number of Feynman diagrams contributing to an $p$-point function
grows factorially, i.e.~super-exponentially, with the order of the coupling constant.
Of course, this is reproduced in the Nicolai-map approach, but it only happens in the final step
of performing the free-field correlation of the composite operator~$X[T_g^{-1}\phi]$.
To see this, consider the $p$-point function in our quantum-mechanical model, 
i.e.~take $X[x]=x(t_1)\,x(t_2)\ldots x(t_p)$.
In the perturbative expansion of $X[T_g^{-1}\,x]$ at order~$g^n$ we find a collection of composite 
operators, which are homogeneous of degree~$p{+}n$ in~$x$. The free correlator then simply applies
Wick's theorem to them, producing all possible free two-point contractions of the $p{+}n$~leaves
shared by the $p$~trees. (An odd number of leaves yields a vanishing correlator of course.)
At this stage we generate also loops (but no purely fermion ones),
and renormalization may be required as usual. The number of such contractions is simply
\begin{equation}
(p{+}n{-}1)!! \= 2^{1-\frac{p+n}{2}}\,\frac{(p{+}n{-}1)!}{(\frac{p+n}{2}{-}1)!}
\ \sim\ \sqrt{2}\,\bigl(\tfrac{p+n}{\ep}\bigr)^{\frac{p+n}{2}}
\ \sim\ \sqrt{2}\,\ep^{-\frac{p+n}{2}}\,n^{p/2}\,n^{n/2}
\qquad\textrm{for}\quad n\to\infty\ .
\end{equation}
It is the final $\sqrt{n}^n$ factor which destroys the convergence and renders the series asymptotic.

\noindent
{\bf Conclusions and outlook.\ }
The Nicolai map of supersymmetric theories allows one to split the computation of quantum correlators into two steps. 
The first step, the perturbative construction of the inverse Nicolai map in terms of tree diagrams, is purely classical. 
We showed that, for a sufficiently small coupling, it converges to a mathematically well-defined functional.
Even though we had to resort to some infrared regularization for the norm estimate of individual tree diagrams,
the crucial point for the convergence of the perturbation series was the merely exponential proliferation 
of unlabelled tree diagrams with an increasing number of nodes. This growth rate is a hallmark of the recursive construction, 
which prevents ``long-distance correlations'' along a tree. 
The second step, a free-field correlation, introduces the quantum loops and factorial growth of the
final perturbation expansion of the correlator.
In this way, the interaction characteristics of the theory have been pushed into the classical domain (tree diagrams!).

One may wonder about renormalization in this scheme. It is known, however, that in globally supersymmetric theories
all coupling renormalization factors are suitable powers of the wave-function renormalization~$Z_{\phi}$ in such a way that
the superpotential~$V$ is unrenormalized. Therefore, upon renormalization, the flow operator~$R_g$ picks up the same
$Z_{\phi}$~power as $\frac{\pa}{\pa g}$, and the map $T_g$ takes precisely the same form in the renormalized quantities.
In other words, the original and the Nicolai-transformed field are renormalized by the same factor.
Subtractions or counterterms need only to be introduced in the second step of the correlator computation.

The considerations of this note also apply, mutatis mutandis, to higher-dimensional field theory including gauge theory.
With a few exceptions, however, the presence of gamma-matrix traces complicates the situation there.
These are akin to an additional loop structure (in spinor space) attached to the tree diagrams and 
lead to Lorentz-index contractions between partial derivatives distributed {\it all over\/} the tree.
Their combinatorics produces a factorial growth already for the number of traced-out terms in the Nicolai map construction. 
However, this additional complexity is comparable to the one in ordinary Feynman perutrbation theory.
Since it obstructs the existence of a polynomial Nicolai map,
it would be very nice to find a way to extend our strategy also to the spin degrees of freedom.

\bigskip

\noindent
{\bf Acknowledgment.\ } 
We acknowledge illuminating discussions with Hermann Nicolai.


\end{document}